# POLLUX: a UV spectropolarimeter for the LUVOIR space telescope project

Eduard Muslimov[*a,b], Jean-Claude Bouret[a], Coralie Neiner[c], Arturo Lopez Ariste[d], Marc Ferrari[a], Sebastien Vives[a], Emmanuel Hugot[a], Robert Grange[a], Simona Lombardo[a], Louise Lopes[e], Josiane Costerate[e], Frank Brachet[e]

[a]Aix Marseille Univ, CNRS, CNES, LAM, Marseille, France; [b]Kazan National Research Technical University named after A.N. Tupolev –KAI, 10 K. Marx, Kazan 420111, Russia; [c] LESIA, Observatoire de Paris, PSL Research University, CNRS, Sorbonne Université, Univ. Paris 06, Univ. Paris Diderot, Sorbonne Paris Cité, 5 place Jules Janssen, 92195 Meudon, France ; [d]IRAP - CNRS UMR 5277. 14, Av. E. Belin.31400 Toulouse. France; [e]Centre National d'Études Spatiales, 18 Avenue Edouard Belin, Toulouse, 31400 France.

## ABSTRACT

The present paper describes the current baseline optical design of POLLUX, a high-resolution spectropolarimeter for the future LUVOIR mission. The instrument will operate in the ultraviolet (UV) domain from 90 to 390 nm in both spectropolarimetric and pure spectroscopic modes. The working range is split between 3 channels – far (90-124.5 nm), medium (118.5-195 nm) and near (195-390 nm) UV. Each of the channels is composed of a polarimeter followed by an echelle spectrograph consisting of a classical off-axis paraboloid collimator, echelle grating with a high grooves frequency and a cross-disperser grating operating also as a camera. The latter component integrates some advanced technologies: it is a blazed grating with a complex grooves pattern formed by holographic recording, which is manufactured on a freeform surface. One of the key features underlying the current design is the large spectral length of each order ~6 nm, which allows to record wide spectral lines without any discontinuities. The modelling results show that the optical design will provide the required spectral resolving power higher than R ~ 120,000 over the entire working range for a point source object with angular size of 30 mas. It is also shown that with the 15-m primary mirror of the LUVOIR telescope the instrument will provide an effective collecting area up to 38 569 $cm^2$. Such a performance will allow to perform a number of groundbreaking scientific observations. Finally, the future work and the technological risks of the design are discussed in details.

## 1. INTRODUCTION

The Large Ultraviolet/Optical/Infrared Surveyor (LUVOIR)[1] is one of four large mission concepts, which were selected for consideration by the 2020 Astronomy and Astrophysics Decadal Survey. The current baseline architecture implies building of a large space telescope with an unprecedented 15-m primary mirror providing a huge collecting area. The LUVOIR mission will cover a number of groundbreaking scientific targets. It will perform a search and the characterization of habitable conditions and signs of life on dozens of potentially habitable worlds beyond our Solar System. LUVOIR will have the ability to characterize hundreds of transiting and directly imaged planets and will revolutionize our understanding of all classes of extrasolar planets and the common threads that connect them. When used for observations of the bodies within the Solar System, LUVOIR can provide up to about 25 km imaging resolution in visible light for Jupiter, monitor atmospheric dynamics in the outer planets, provide high-resolution imaging and spectroscopy of Solar System comets, asteroids, moons, and Kuiper Belt objects and perform many other tasks. In the field of cosmology LUVOIR can address such questions as the nature of dark matter and galaxy formation processes on small size scales by imaging fainter and smaller structures in the universe than ever before. From the point of galaxy evolution studies, LUVOIR's unprecedented resolution will resolve stellar populations in star-forming regions of galaxies at distances up to 10-25 Mpc, accessing more diverse galaxy morphologies, sizes, and cluster environments. In addition, the UV capability provided by LUVOIR will be vital for discriminating among our theories of star and planet formation. Finally, the UV spectropolarimetry capability of LUVOIR will allow us to study, e.g., linear polarization and depolarization effects from circumstellar disks and stellar magnetic fields. This list of science goals is not exhaustive, but it clearly shows the huge scientific impact, which the mission may have.

---

[*] eduard.muslimov@lam.fr; phone +33 4 91 05 69 18; lam.fr

According to the current baseline the telescope payload consists of 4 instruments. In the current paper we consider the optical design of high-resolution UV spectropolarimeter called POLLUX, which represents the European contribution to the mission.

## 2. BASELINE OPTICAL ARCHITECTURE

### 1.1. Top-level requirements

Based on the key science goals of the mission the science groups of the POLLUX consortium have defined the following essential top-level requirements for the instrument (see Table1).

Table 1. High-level requirements to the spectropolarimeter.

| Parameter | Requirement | Goal |
|---|---|---|
| **Shortest wavelength** | 90 nm | 90 nm |
| **Longest wavelength** | 390 nm | visible |
| **Spectral resolving power** | 120 000 | 200 0000 |
| **Spectral length of the order** | 6 nm | ≥6 nm |
| **Polarisation mode** | Circular+linear (= QUV) | |
| **Aperture size** | 0.03" | 0.01" |
| **Observing modes** | with and without polarimetry, i.e. spectropolarimetric and pure spectroscopic modes | |

The only type of spectrograph optical design that allows to reach the target high spectral resolving power is an echelle spectrograph[2]. Such a spectrograph consists of collimating and focusing optics and two dispersing elements. The main disperser is an echelle grating – a grating with relatively low spatial frequency, working with a high incidence angle in a large number of higher diffraction orders. It forms the corresponding number of overlapped spectral images stretched in the X direction. The second disperser or cross-disperser is an element which separates the orders of the echelle grating. Its dispersion operates along the perpendicular Y axis.

### 1.2. Basic design solutions

The instrument optical design architecture is based on the following assumptions about the components and technologies. We must note that most of the points below are provided by the LUVOIR project.

1) We assume that the telescope is a three-mirror anastigmate (TMA) with a collecting area of 139-173.4 m$^2$ and it operates with an *f*-ratio of 20.48. We also assume the instrument picks off the beam close to the field of view center – the field X and Y coordinates are (0.015°, -0.020°). The telescope residual aberrations are neglected.

2) It is assumed that a pinhole is placed at the instrument entrance. When a pinhole is used instead of a slit, a simpler aberration correction conditions may be used and all the channels can be rotated around the chief ray. The latter feature makes the design more flexible.

3) The working spectral range should be split into sub-ranges, because it exceeds one full octave. Such an approach also allows to achieve high spectral resolving power with feasible values of the detector length, the camera optics field of view and the overall size of the instrument. In addition, it allows to use dedicated optical elements, coatings and detectors for each band and obtain a gain in efficiency. The entire spectral range is thus divided into far ultraviolet (FUV), medium ultraviolet (MUV) and near ultraviolet (NUV). Each of the bands is recorded by a dedicated channel.

4) The MUV and NUV channel are separated by means of a dichroic splitter. It was demonstrated that such a splitter can have a high efficiency[3]. Also, in comparison with other separation techniques such as a spatial splitter or a rotating mirror, a dichroic splitter allows to work in two bands simultaneously and use the full aperture thus achieving the high resolving power with relatively a small collimator focal length.

5) Due to the channels separation method, chosen above, it is impossible and unnecessary to provide a considerable overlap between the MUV and NUV bands. It is assumed to be a sharp boundary. The longwave limit for the NUV is the same as that for the entire instrument, i.e. 390 nm. Thus, to have one full octave in the NUV we set the MUV/NUV boundary at 195 nm.

6) Currently there are no dichroic splitters operating in the FUV below the Lyα line and there is no data that such an element will become possible in the future. So we propose to use a flip mirror to feed the FUV channel. It should be located right before the dichroic splitter, because the dichroic substrate cannot transmit the FUV irradiation.

7) Accounting for the chosen FUV splitting method, the FUV and MUV boundaries are set as follows. One of the most important spectral lines in the UV is the Lyα line centered at ~121.5 nm. Depending on the conditions its width can reach several nm. Therefore, the shortest limit for the MUV band is set at 118.5 nm, while the longest one for the FUV is 124.5 nm. Thus, the Lyα line appears in both channels.

8) The shortest wavelength for the FUV is the same as that for the entire instrument and equals to 90 nm. This value strongly depends on the main telescope throughput and may be reconsidered in the future.

9) In each channel the beam is collimated by an ordinary off-axis parabolic (OAP) mirror. The off-axis shift and the corresponding ray deviation angle are chosen in such a way that the distance between the entrance pinhole and the echelle is large enough to place the polarimeter and corresponding mechanical parts etc.

10) Echelle grating works in a quasi-Littrow mounting. The exact values of the grooves frequency and the blazing angle are computed to obtain the required dispersion and subsequently the spectral resolving power. No boundary conditions were directly applied to these computations.

11) The cross-disperser in each channel operates also as a camera mirror, so it is a concave reflection grating. This approach allows to decrease the number of optical components and increase the throughput. In order to correct the aberrations we suppose that the cross-disperser's surface is a freeform and it has a complex pattern of grooves formed by holographic recording (i.e. interferometric technique using two coherent point sources).

12) The detector is represented only by dimensions of its sensitive area and the pixel size. The baseline configuration uses δ-doped CCD's in all channels. The achievable parameters were recently reported in a number of publications[4,5].

13) Polarimeters are placed right after the splitters in each channel. The polarimeters should be retractable at least in the MUV and NUV channels. Such a solution will provide a considerable gain of throughput in the pure spectroscopic mode. For the FUV polarimeter only the modulator part is removable.

14) The change of optical path caused by the polarimeter removal is compensated by means of a change of the OAP mirror.

15) The polarimeter principle design was chosen separately for each channel accounting for the technological feasibility and the expected performance: a birefringent modulator and a Wollaston prism for the NUV (similar to the design for the Arago instrument[6]; see also Le Gal et al., these proceedings); a three-mirror modulator and a Wollaston prism for the MUV; and a three-mirror modulator and a Brewster angle analyzer for the FUV.

Accounting for all the assumptions and basic design solutions listed in the previous paragraph we propose the baseline optical architecture shown in Fig.1 (note that the linear and angular values on this schematic drawing are not to scale).

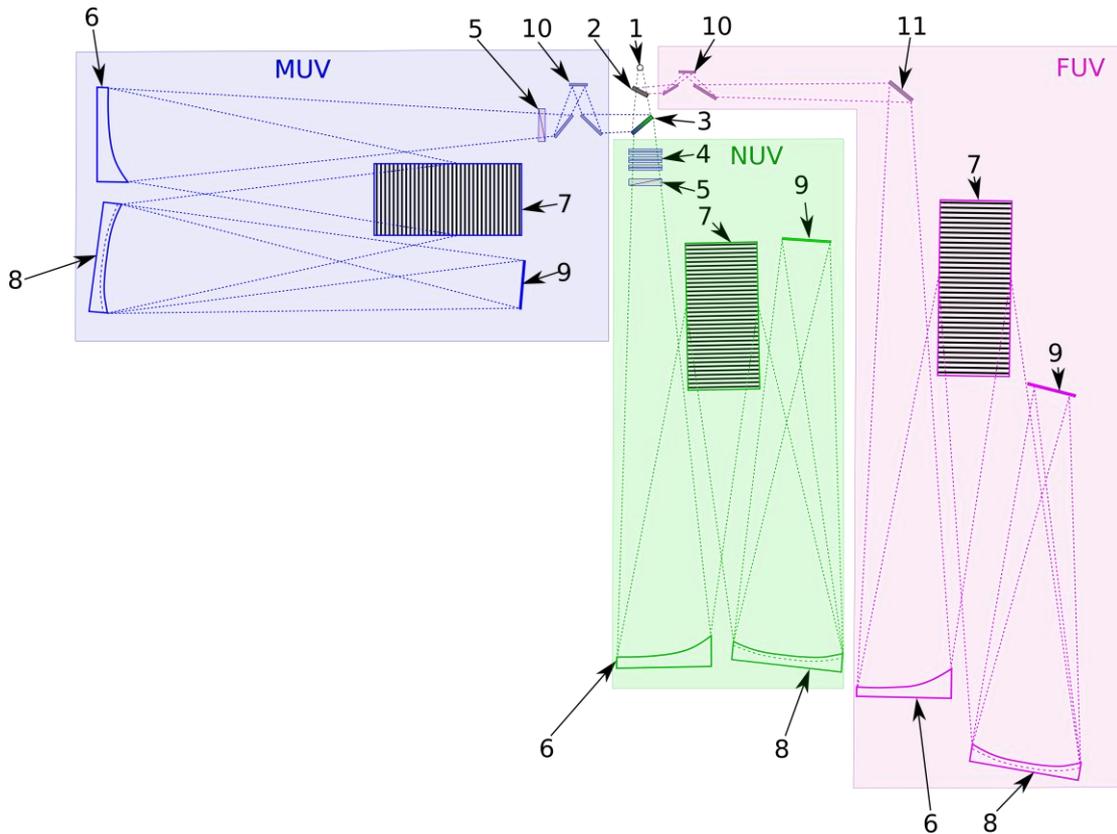

Figure 1. POLLUX instrument baseline architecture schematic diagram: 1- pinhole, 2-flip mirror, 3-dichroic splitter, 4-birefringent modulator, 5-Wollaston analyzer, 6-OAP collimator (2 mirrors for different operational modes, replaceable by translational mechanism), 7- echelle grating, 8- concave freeform holographic grating, 9-detector array, 10- 3-mirror modulator, 11- Brewster angle analyzer.

## 3. OPTICAL DESIGN

For the development of the optical design with real optical components we made two additional assumptions. First, in order to account for possible misalignments and manufacturing errors as well as for the resolution changes due to switching from the pure spectroscopic mode to the spectropolarimetric one, the target spectral resolving power was set to 130 000 (i.e. 110% of the requirement). Second, the MUV and NUV optical schemes were unified as much as possible: they have identical collimators, the same camera focal length, the same detector format and similar echelle blazing angles. It will simplify the manufacturing, optical testing and assembling.

### 1.3. Optical design overview

The optical design of each channel was optimized according to the conditions described above. The polarimeter units were integrated into the same models. The optimization was performed with the standard tools provided by the Zemax software. In order to account for different wavelengths and polarization states 24 configurations representing 4 orders were used. The spot RMS size was used as the main image quality criterion with the weight coefficient of 5 assigned to its X size. The boundary conditions were used to control the image centering, the image lines (i.e. orders and polarization components) separation, the air gap between the echelle grating and detector and the arms lengths in the holographic grating recording geometry. In addition, for the MUV and NUV designs the Wollaston prisms wedges edge thicknesses are constrained.

The MUV channel optical design general view is shown as an example in Fig.2 (note that the rays propagation direction differs from the actual one because of the software conventions). The NUV and FUV channels repeat the same principal geometry, though the polarization units are different. The key design solutions made for each of the components are described in the next sections.

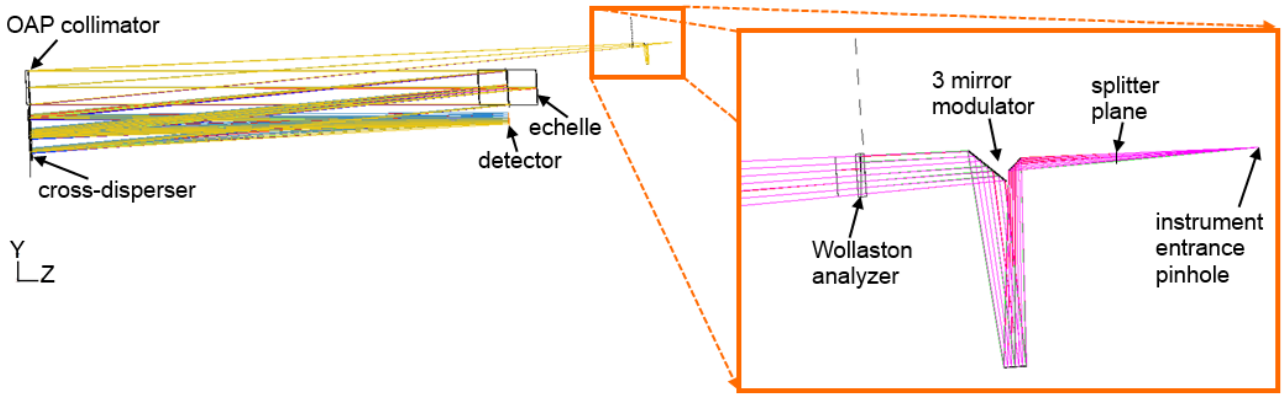

Figure 2. MUV channel optical design: left – general view, right – zoom-in of the polarimetric unit.

### 1.4. Optical system components

The qualitative description for the optical components design is given below in the order of radiation propagation:

1) The splitters are placed as close to the focal point as possible in order to decrease their size: the flip mirror is located 20 mm from the focus and the distance from the focus to the dichroic is 35 mm.

2) The polarimeters should have the minimal size in order to decrease the defocusing and chromatic aberrations and simplify the observation mode switching. The NUV polarimeter is inspired by the Arago design[6] (see Le Gal et al., these proceedings).. The final design of the MUV and FUV polarimeters will depend on the exact materials properties. In the pure spectroscopic mode the MUV and NUV polarimeters are removed from the optical path; the FUV polarimeter is static.

3) All the collimators are usual OAP mirrors. The collimators may introduce additional aberrations because of the polarization components separation. For the defocus compensation when switching from the spectropolarimatric to the spectroscopic mode, the OAP is replaced by another one with a translation platform.

4) The MUV and NUV echelle gratings have similar sizes and blazing angles as well as the frequencies that differ by a factor of ~2. All of these parameters values are achievable with the current technology level[7]. The FUV echelle grating is unique because of its frequency and blazing angle.

5) Each of the cross-dispersers is a concave freeform holographic grating. It is a novel type of optical elements having high aberration correction capabilities. The surface shape is described by a vertex sphere and 5 Zernike terms corresponding to the main aberrations. Such a surface has no axial symmetry, but it is still symmetrical with respect to the YZ plane. The freeform complexity is assessed by deviation from the best fit sphere (BFS). For all the cross-dispersers this deviation is relatively small. The grooves are formed by recording of an interference pattern from two coherent point sources. A sufficient aberration correction is possible only if the spectral components at the cross-disperser's surface are separated. However, it leads to the grating's aperture growth.

6) In the current study the detector is represented as an array of ideal sensitive elements. As it was mentioned before, the MUV and NUV detectors have the same geometry.

The quantitative description of the optical schemes components is given below in Table 2.

Table 2. Design parameters of the optical system components.

| Channel | NUV | MUV | FUV |
|---|---|---|---|
| **Wavelength range** | 195-390 nm | 118.5-195 nm | 90-124.5 nm |
| **Polarimeter** | | | |
| **Modulator** | 3 pairs of MgF$_2$ plates | 3 mirrors, the first incidence angle=47º | 3 mirrors, the first incidence angle=80º |
| **Thicknesses, mm** | 0.8 mm thickness, 0.2 mm separation | 50 (x2) | 20 (x2) |
| **Clear aperture max, mm** | 4.4 | 8x12 | 4.2x23 |
| **Analyzer** | MgF$_2$ Wollaston prism | MgF$_2$ Wollaston prism | SiC plate at the Brewster angle |
| **Prism angle, º** | 5.027 | 3.175 | 33.365 |
| **Clear aperture max, mm** | 6.6 | 9.6 | 5.5x6.6 |
| **Wedge thickness center/edge, mm** | 1/0.716 | 1/0.74 | N/A |
| **Rays separation at detector, µm** | 200 | 200 | N/A |
| **Collimator** | | | |
| **Focal length, mm** | 1702.692 | 1702.692 | 3109.396 |
| **Clear aperture, mm** | 88 | 87 | 154.5 |
| **Decenter, mm** | 120.2 | 120.2 | 210 |
| **Deviation angle, º** | 4.041 | 4.041 | 3.873 |
| **Echelle** | | | |
| **Frequency, 1/mm** | 144.95 | 289.4 | 589.7 |
| **Angle, º** | 59.551 | 59.388 | 42.779 |
| **Clear aperture XxY, mm** | 89.2x168.6 | 87.4x167.4 | 154.7x210.6 |
| **Orders** | 31-61 | 31-50 | 19-26 |
| **Order spectral length, nm** | 3.2-12.2 | 2.4-6.1 | 3.4-6.3 |
| **Typical orders separation, mm** | 0.7 | 0.9 | 2.6 |
| **Cross-disperser** | | | |
| **Focal length, mm** | 1200 | 1200 | 2100 |
| **Clear aperture XxY, mm** | 216x99.5 | 215.4x98.3 | 416.1x165.4 |
| **Asphericity RMS/PTV, µm** | 2.15/3.21 | 2.27/3.36 | 2.4/3.7 |
| **Frequency, 1/mm** | 105.4 | 212.3 | 272.6 |
| **Rec. sources coordinates, (y,z) mm** | (30.399,1614.151) and (-53.760,1648.412) | (100.194,1913.946) and (-99.558,1936.898) | (163.034,2534.301) and (-179.888,2606.124) |
| **Detector** | | | |
| **Type** | δ-doped CCD | δ-doped CCD | δ-doped CCD |
| **Size XxY, mm** | 131x24 | 131x19 | 202.7x19.4 |
| **Pixel size, µm** | 13 | 13 | 13 |
| **Sampling** | 2.4 | 2.4 | 2.3 |

# 4. PERFORMANCE

### 1.5. Image quality and spectral resolution

During the optical design development and optimization the image quality is controlled through spot diagrams. They are used for the image centering on the detector and maintain the orders and polarization components separation as well as for the aberrations minimizations. An example of full-frame spot diagram for the MUV channel is given in Fig. 3. It clearly shows the detector filling and the lines separation.

A typical spot diagram for an individual configuration is shown in Fig.4 (right part). Since the resolution is defined only by the line spread function along the X axis, the spot Y size had ~5-time lower weight coefficients during the numerical optimization. Therefore, the spot diagrams are elongated along the Y axis.

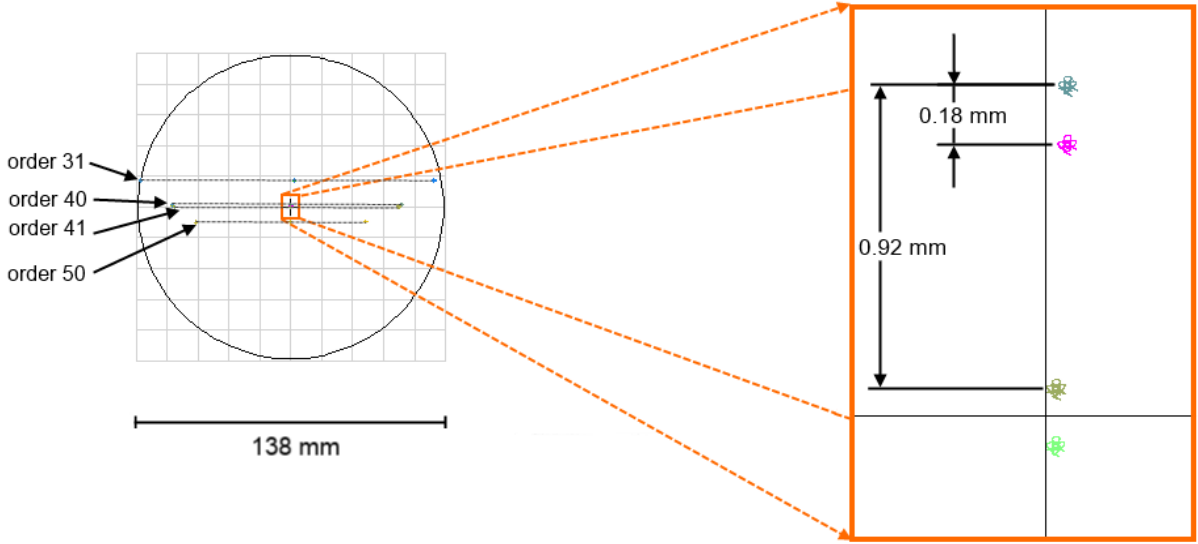

Figure 3. The MUV channel image: left – the detector filling diagram, right – the zoom-in showing the orders separation and the polarization components separation in each order.

The final control on the image quality and the spectral resolution computation were performed with use of instrument functions (IF). The IF in this case is a relative distribution of illumination in the pinhole monochromatic image measured along in the XZ section. It is computed as a convolution of the optical system line spread function and the pinhole rectangular function. It was assumed that the spectral resolution in wavelength units equals to

$$\Delta\lambda = \Delta y' \cdot \frac{\partial \lambda}{\partial L}, \qquad (1)$$

where $\Delta y'$ is the instrument function full width at the half maximum (FWHM) and $\partial\lambda/\partial L$ is the reciprocal linear dispersion. A typical IF is shown in Fig.4 (left part).

The spectral resolving power is defined as

$$R = \frac{\lambda}{\Delta\lambda}, \qquad (2)$$

where $\lambda$ is the wavelength.

At the same time the theoretical limit is given by[8]

$$R = \frac{2D_{col} tg\varphi}{D_{TE} tg\delta} \Delta\lambda, \qquad (3)$$

Here $D_{col}$ is the collimated beam diameter, $\varphi$ is the echelle blazing angle, $D_{TE}$ is the effective diameter of the telescope primary mirror, $\delta$ is the angular size of the entrance slit or pinhole. The actual resolution is defined by the smallest value between the ones obtained from (2) and (3).

The diagram showing the results of spectral resolving power computations for all the channels is given in Fig.5. The separate datapoints correspond to the values computed with Equations (1)-(2) using the aberrated IFs. The horizontal lines shows the values obtained with Equation (3) as well as the target value (see Table 1) and the corrected target value accounting for possible misalignments and the mode change (see Section 3). As one can see, the spectral resolution requirements are fulfilled for all wavelengths. There is only one exclusion at the edge of the NUV band. This resolution loss appears because of the polarization ray splitting and clearly demonstrates the optical design sensitivity.

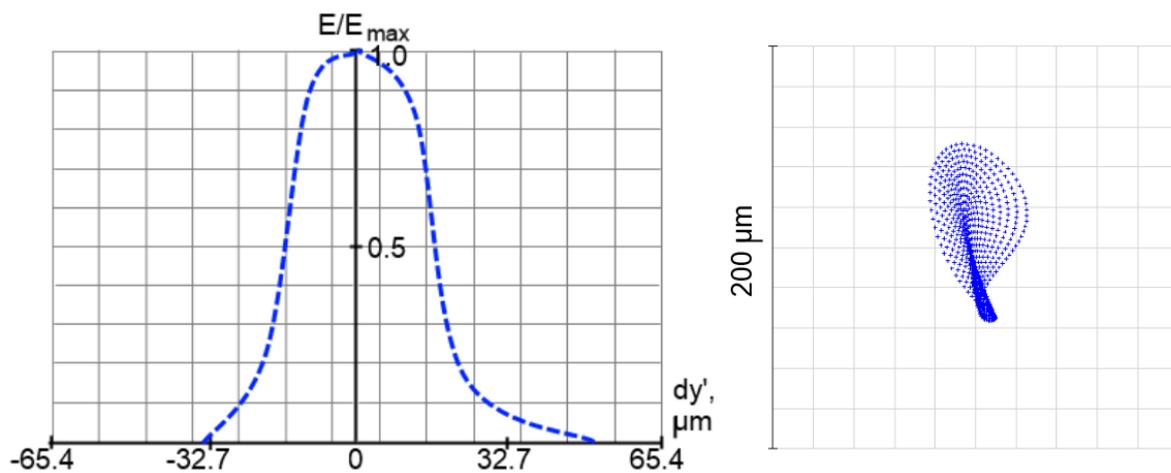

Figure 4. Typical optical quality indicators at 377.8 nm: left – instrument function (FWHM is 31.2 μm), right – spot diagram (RMS radius is 25.1 μm).

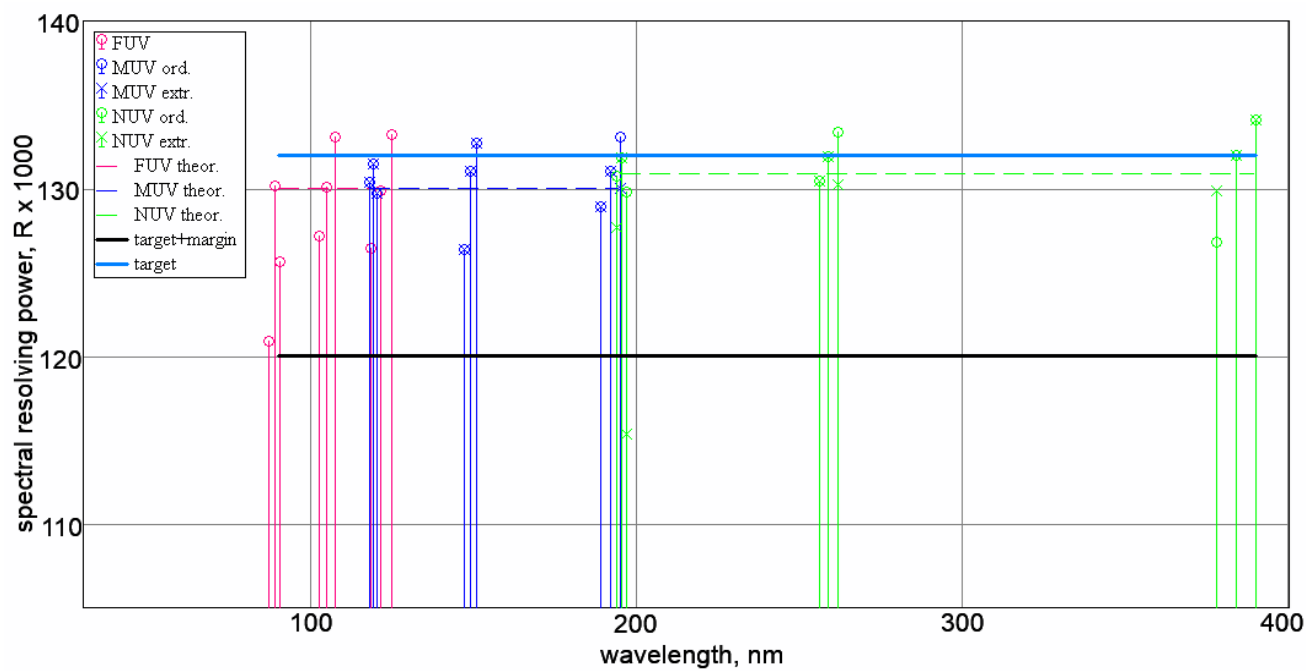

Figure 5. Spectral resolving power diagram for the POLLUX channels. The theoretical limits are: FUV – 129979, MUV– 130008, NUV– 130855.

### 1.6. Throughput

Below we consider the efficiency of each element type point-by-point along the optical train. The data used for this computations was taken from the literature or, when possible, found with simplified simulations. We must note that in order to avoid issues with sampling difference, noises and high frequency irregularities in the initial data was interpolated. So all the spectral dependences are presented by smoothened envelope curves.

1) The pick-off mirror has to be efficient for all channels (MUV, NUV, and FUV). Therefore, a coating with a SiC capping layer on a metal base layer will be required, namely – Al+MgF$_2$+SiC. It will provide 30% to 40% in the FUV channel and higher values in the NUV and MUV channels[9].

2) For the dichroic splitter characteristics, we used the data from the GALEX splitter[3]. For simplicity we assumed that the throughput curves shapes remain the same, but they are shifted to fit with the splitting border of 195 nm (instead of 178 nm for GALEX).

3) The current assumptions concerning the reflective coatings are as follows. For the reflective elements in the FUV a single layer of SiC will be used. Its reflection can exceed 40% in a nearly-normal incidence[10]. The baseline conception also assumes that the rest of the mirrors use the same coating as the main LUVOIR telescope. It will be, most likely, a multilayer coating consisting of Al+LiF+AlF$_3$. It has high reflectance in the NUV and MUV; also it is notable for its stability[11,12].

4) The birefringent material used in the MUV and NUV polarimeters is MgF$_2$. Its properties are described in a number of sources[13].

5) The three-mirror modulators in the FUV and MUV operate with high incidence angles. Therefore, their efficiencies were computed separately. It was supposed here that the polarimeter's mirrors for the MUV are coated by Al+LiF and those for the FUV are with a SiC layer, which behaves as a dielectric. Also, it is supposed that the Brewster angle analyzer in the FUV is completely lossless.

6) The customized echelle gratings can be produced on a Si substrate by means of photo-lithography and *e*-beam-etching[7,14]. Due to the material properties the peak of the triangular groove profile has an angle of 70.6º instead of 90º. The echelle gratings coating is 17 nm of LiF on Al. In addition, for the efficiency estimation the deviation from the Littrow mounting was neglected. We must emphasize that at the current stage possible signal summation for the overlapping parts of adjacent orders is not accounted for. Thus the minimum efficiency may be increased, though the maxima will remain the same.

7) The cross-dispersers grooves efficiency was computed with use of the RCWA (rigorous coupled-wave analysis) method, implemented in the *GD-calc* software[15]. The computation was performed for all the control wavelengths and 2 polarization states over a 3x3 probe grid covering the pupil. It accounted for the actual ray direction and the freefrom surface normal, but the local changes of the grooves frequency and curvature were neglected. It was supposed that the groove profile is triangular blazed for the central wavelength for each sub-range and the profile depth is ½ of this wavelength. Below average values computed for the 9 rays and for 2 polarizations are used.

8) All the detectors in the baseline design are δ-doped CCD. They have relatively high quantum efficiency down to the extreme UV[4,5]. Here we apply the QE data for an uncoated CCD (otherwise, the efficiency in the FUV will be significantly decreased) and use an envelope curve.

9) Finally, we account for the reflection losses inside the telescope. It was assumed that the telescope mirrors are coated by the Al+LiF+AlF$_3$ coating and the working beam experiences 4 bounces with a small angle of incidence [1].

The overall throughput was found as the product of all the elements efficiencies in the corresponding channels. Then it was converted to the equivalent effective area by multiplication by a geometrical factor[16] equal to $A_{geom}$=135 000 cm$^2$. It allows to account for the LUVOIR's large collecting area. Two cases were considered – coupling with the current telescope and coupling with an ideal 100%-transmitting telescope. The latter computation is necessary to assess the performance of POLLUX itself. The results are shown in Fig.6.

Even with the relatively low throughput the instrument can show a high performance due to the huge collecting area of the telescope. The maximum values of the effective area are 546, 3917 and 38569 cm$^2$ for the FUV, MUV and NUV channels respectively in the cases when all the losses inside the telescope and polarimetric unit are accounted for. These values should be sufficient for the most of our scientific goals. One can note that the FUV throughput drops to zero below 103 nm. This is explained by the telescope's cut-off and represents a point of ongoing discussion with the LUVOIR team and future improvement.

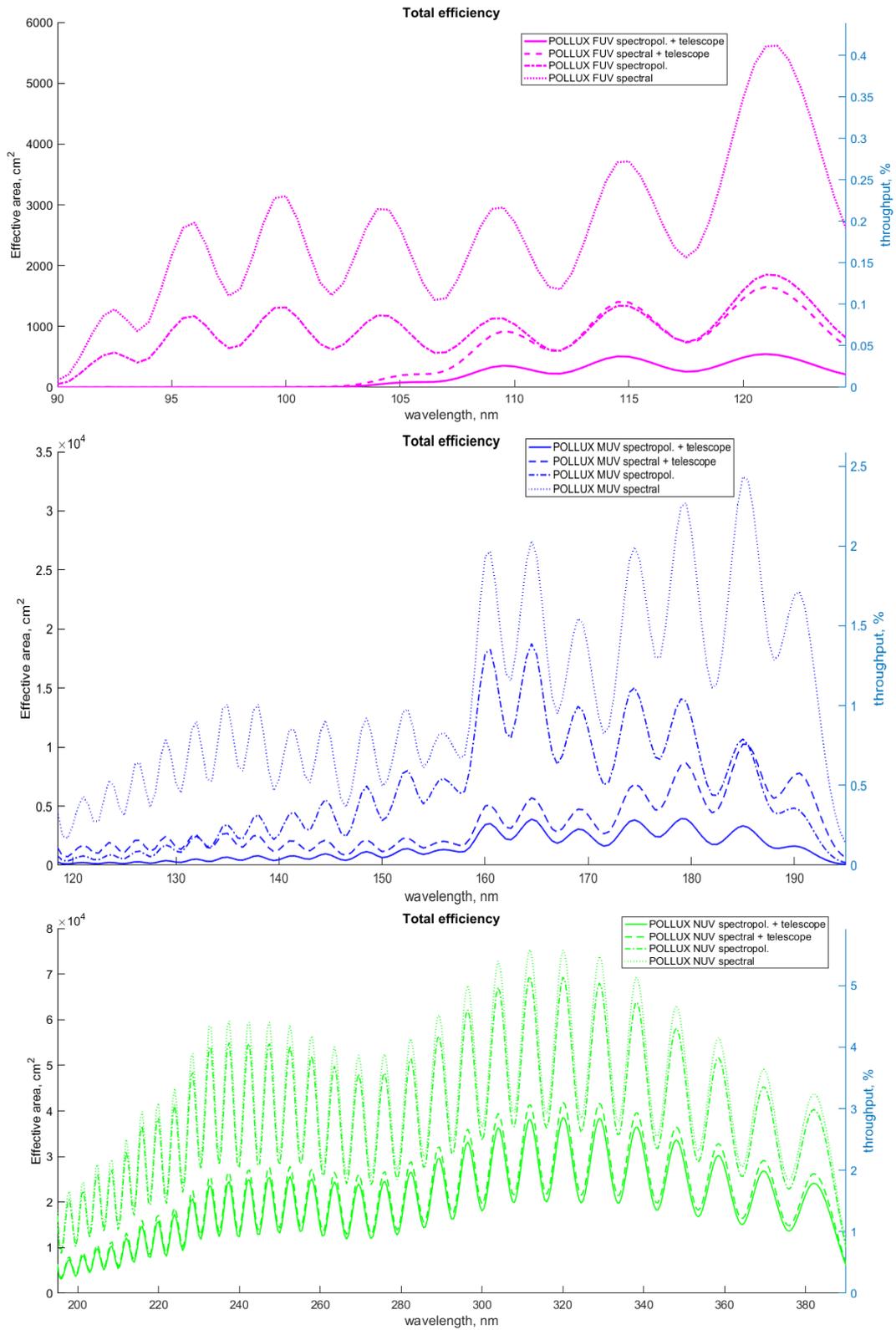

Figure 6. Overall efficiencies and the corresponding effective areas for the POLLUX channels, top-to-bottom – FUV, MUV, NUV.

# 5. COUPLING WITH THE TELESCOPE

POLLUX will be mounted behind the LUVOIR's deployable primary mirror in one of the 4 dedicated volumes (see Fig.7). The pick-off mirror directs a portion of the telescope beam to the instrument entrance. The angular coordinates of the entrance pinhole projection to the sky are (-0.015º, -0.020º).

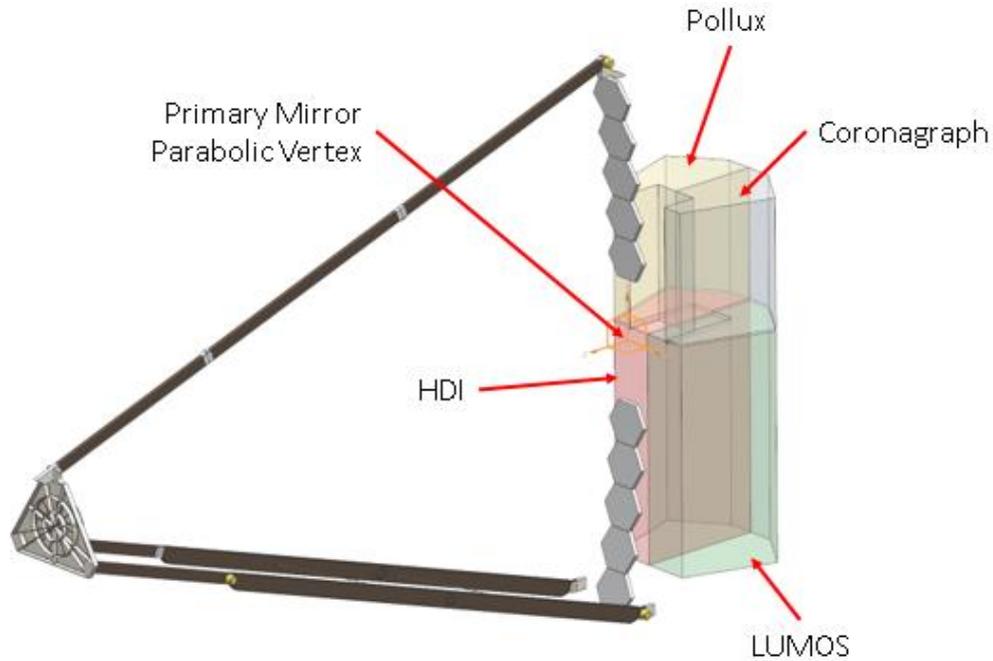

Figure 7. Location of the POLLUX dedicated volume with respect to the telescope and other payload instruments. The maximum dimensions of the volume are 3600x2147x3688 mm$^3$ (LxWxH) (courtesy of the Goddard Space Flight Center).

Fig.8. shows the arrangement of channels and optical components inside the volume. This scheme will be used as a starting point for the opto-mechanical design.

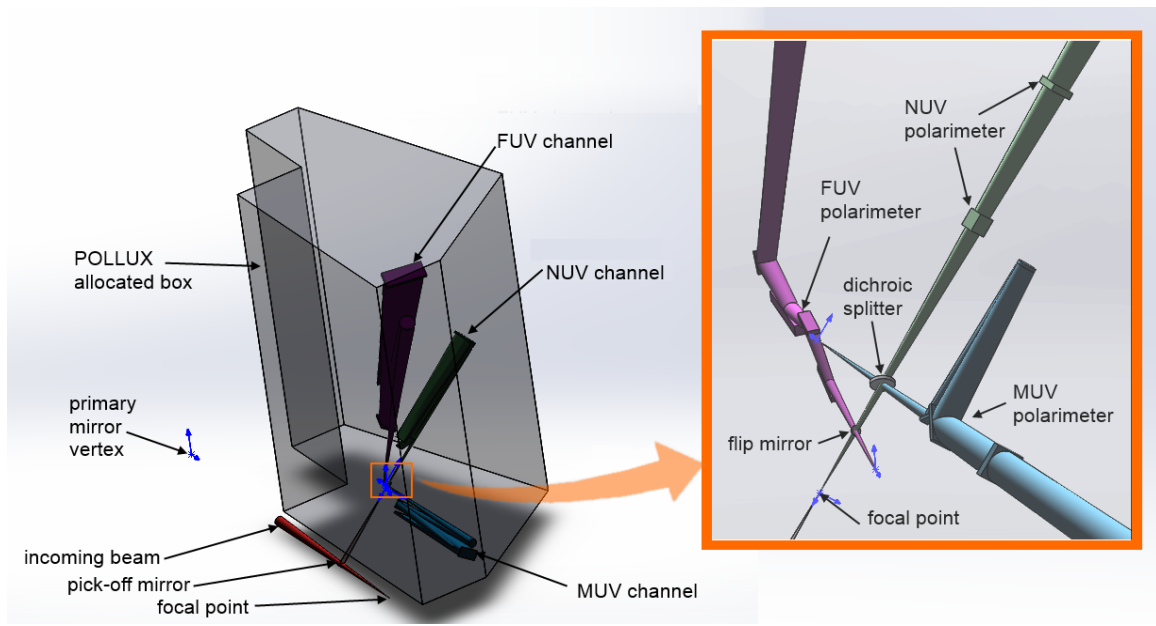

Figure 8. Optical system of POLLUX arranged inside the dedicated volume.

## 6. CONCLUSIONS AND FUTURE WORK

In the present paper we presented the preliminary design concept of the POLLUX instrument, a high-resolution UV spectropolarimeter for the LUVOIR space mission project. Thanks to the recent developments in optical components and detectors technologies it will be able to reach the target performance and cover a number of groundbreaking science cases.

During this preliminary study a number of technological risks and critical points, which define the future work, were found. Below we provide a short overview of them:

1) There is a risk related to the reflective coatings properties ambiguity. All the coatings, especially the ones used in the FUV channel should be examined under different conditions, including incidence angles and polarization states and be space-qualified. Also, the deposition technique[17] and alternative materials should be considered in details[18].

2) It was assumed that the dichroic splitter has a performance similar to that of the GALEX dichroic. In the future, the reflectivity and transmission as well as the working angles of the POLLUX dichroic should be studied separately. Moreover, the FUV channel will be placed in the direct propagation after the flip mirror, while the MUV/NUV will be placed in reflection in order to increase the FUV throughput.

3) The size of the FUV echelle grating exceeds the maximum size of a grating ever produced with the chosen technology. So this is a technological risk and a subject for future studies.

4) The possibility to fabricate a holographic freeform grating with triangular grooves must be demonstrated in practice. Scaling of such an element represents a separate technological challenge. We should note that the backup solution for this element is a variable line spacing ruled freeform grating.

5) The detector parameters are also subject to a further investigation. Use of CMOS instead of CCD is an option. Because the detectors have large dimensions, we should study options of tiling. Furthermore, in the future one may consider the detector's anti-reflective coating properties.

6) Finally, a number of necessary analyses have not been performed yet including tolerance analysis and ghost analysis.

## ACKNOWLEDGEMENTS

The authors acknowledge the entire LUVOIR and POLLUX consortia for their contributions. We would like to especially thank Kevin France (Univ. of Colorado – Boulder), Matthew Bolcar (NASA GSFC), Aki Roberge (NASA GSFC), Randall McEntaffer (Penn State Univ.), Chris Evans and David Montgomery (UK ATC). Eduard Muslimov acknowledges the support from the European Research council through the H2020 - ERCSTG-2015 - 678777 ICARUS program.

## REFERENCES

[1] M. R. Bolcar et al., "The Large UV/Optical/Infrared Surveyor (LUVOIR): Decadal Mission concept design update," Proc. SPIE 10398, 1039809 (2017)
[2] T. Eversberg and K. Vollmann,[Spectroscopic Instrumentation: Fundamentals and Guidelines for Astronomers], Springer, Heidelberg (2015)
[3] GALEX: Chapter 1 – Instrument overview, http://www.galex.caltech.edu/researcher/techdoc-ch1.html
[4] Nikzad S. et al., "High-efficiency UV/optical/NIR detectors for large aperture telescopes and UV explorer missions: development of and field observations with delta-doped arrays," J. Astron. Telesc. Instrum. Syst. 3(3), 036002 (2017)

[5] Nikzad, S. at al., "Single Photon Counting UV Solar-Blind Detectors Using Silicon and III-Nitride Materials," Sensors 16, 927 (2016)
[6] M. Pertenais et al. "Optical design of Arago's spectropolarimeter," Proc. SPIE 10562, 105622A (2017)
[7] McEntaffer, R., DeRoo, C., Schultz, T. et al. ,"First results from a next-generation off-plane X-ray diffraction grating", Exp Astron (2013)
[8] C.R. Kitchin, [Optical Astronomical Spectroscopy] Taylor&Francis, New York (1995)
[9] J. I. Larruquert and R. A. M. Keski-Kuha, "Multilayer coatings with high reflectance in the extreme-ultraviolet spectral range of 50 to 121.6 nm," Appl. Opt. 38(7), 1231 (1999)
[10] M. Fernández-Perea , J. I. Larruquert, J. A. Aznárez, J. A. Méndez, "Transmittance and reflective coatings for the 50-200 nm spectral range," Proc. SPIE 631 (2006)
[11] M. R. Bolcar, "UV Coatings and Short-wavelength Cutoff ," LUVOIR Tech Note Series (2016)
[12] K. Balasubramanianaet al., "Aluminum Mirror Coatings for UVOIR Telescope Optics including the Far UV," Proc. SPIE 9602, 96020I (2015)
[13] M. Wakaki,T. Shibuya,K. Kudo, [Physical Properties and Data of Optical Materials], CRC Press, Boca Raton (2007)
[14] H. Marlowe et al., "Modeling and empirical characterization of the polarization response of off-plane reflection gratings," Appl. Opt. 55, 5548-5553 (2016)
[15] K.C. Johnson ,Grating Diffraction Calculator (GD-Calc® ) – Demo and Tutorial Guide (2006) http://kjinnovation.com/GD-Calc_Demo.pdf
[16] K. France et al., "The LUVOIR Ultraviolet Multi-Object Spectrograph (LUMOS): instrument definition and design," Proc. SPIE 10397, 1039713 (2017)
[17] A. Fludra, et al., "SPICE EUV spectrometer for the Solar Orbiter mission," Proc. SPIE 8862, 88620F (2013)
[18] J. Del Hoyo and M. Quijada, "Enhanced aluminum reflecting and solar-blind filter coatings for the far-ultraviolet," Proc. SPIE 10372, 1037204 (2017)